\documentclass[prl,twocolumn,amsmath,amssymb,superscriptaddress]{revtex4-1}
\usepackage{amsfonts}
\usepackage{amsmath}
\usepackage{amssymb}
\usepackage{graphicx}
\usepackage[T1]{fontenc}
\usepackage[utf8]{inputenc}
\usepackage{hyperref}
\usepackage[dvipsnames]{xcolor}
\usepackage{color}
\usepackage{bm}
\usepackage{comment}
\usepackage{mathtools}
\usepackage[section]{placeins}
\usepackage[capitalize]{cleveref}
\usepackage{footnote}
\usepackage{etoolbox}
\usepackage[caption=false]{subfig}

\usepackage{pgfplots} 
\usetikzlibrary{plotmarks}
\pgfplotsset{compat=newest} 
\pgfplotsset{plot coordinates/math parser=false}

\crefrangelabelformat{equation}{(#3#1#4--#5\crefstripprefix{#1}{#2}#6)}

\begin{document}
\title{Experimental measurement of the isolated magnetic susceptibility}

\author{D. Billington} 
\affiliation{School of Physics and Astronomy, Cardiff University, CF24 3AA Cardiff, United Kingdom}

\author{C. Paulsen}
\affiliation{Institut N\'eel, CNRS $\&$ Universit\'e Grenoble Alpes, 38000 Grenoble, France}

\author{E. Lhotel}
\affiliation{Institut N\'eel, CNRS $\&$ Universit\'e Grenoble Alpes, 38000 Grenoble, France}

\author{J. Cannon} 
\affiliation{School of Physics and Astronomy, Cardiff University, CF24 3AA Cardiff, United Kingdom}

\author{E. Riordan} 
\affiliation{School of Physics and Astronomy, Cardiff University, CF24 3AA Cardiff, United Kingdom}

\author{M. Salman} 
\affiliation{School of Physics and Astronomy, Cardiff University, CF24 3AA Cardiff, United Kingdom}

\author{G. Klemencic} 
\affiliation{School of Physics and Astronomy, Cardiff University, CF24 3AA Cardiff, United Kingdom}

\author{C. Cafolla-Ward} 
\affiliation{School of Physics and Astronomy, Cardiff University, CF24 3AA Cardiff, United Kingdom}

\author{D. Prabhakaran}
\affiliation{Clarendon Laboratory, Physics Department, Oxford University,Oxford, OX1~3PU, United Kingdom}

\author{S. R. Giblin} 
\affiliation{School of Physics and Astronomy, Cardiff University, CF24 3AA Cardiff, United Kingdom}

\author{S. T. Bramwell}
\affiliation{London Centre for Nanotechnology and Department of Physics and Astronomy, University College London, 17-19 Gordon Street, London, WC1H OAH, U.K.}

\begin{abstract}

The isolated susceptibility $\chi_{\rm I}$ may be defined as a (non-thermodynamic) average over the canonical ensemble, but while it has often been discussed in the literature, it has not been clearly measured. Here, we demonstrate an unambiguous measurement of $\chi_{\rm I}$ at avoided nuclear-electronic level crossings in a dilute spin ice system, containing well-separated holmium ions. We show that $\chi_{\rm I}$ quantifies the superposition of quasi-classical spin states at these points, and is a direct measure of state concurrence and populations.

\end{abstract}

\maketitle

\section{I Introduction}

Alternating current susceptometry~\cite{Casimir1938,Topping} is a traditional probe of magnetic response at applied frequencies $\omega/(2\pi)$ of up to $10^6$ Hz~\cite{Riordan2019}. As $\omega \rightarrow 0$, the isothermal susceptibilty $\chi_T$ is measured, while historically there was much debate as to whether the high frequency response could be approximated as a quasi-static adiabatic susceptibility, $\chi_S$~\cite{Casimir1938}, or a quasi-static isolated (or quantum adiabatic) susceptibility, $\chi_{\rm I}$~\cite{Broer1951,Wilcox1968,Pirc1974}. The latter is a particularly interesting response function as it reveals aspects of a system that are not exposed by thermodynamic measurements, yet there do not seem to be any examples where $\chi_{\rm I}$ has been clearly observed~\footnote{Except perhaps by Amaya et al. in the context of cross relaxation: K. Amaya, Y. Tokunaga, Y. Kuramitsu and T. Haseda, J. Phys. Soc. Japan {\bf 33}, 49 (1972).}. Here, we demonstrate an experimental measurement of $\chi_{\rm I}$ at avoided level crossings in a simple spin system, and show how it is a direct measure of the concurrence, or superposition of two quasi-classical spin states, and can be used to measure state populations.

The three susceptibilities, $\chi_{T}$, $\chi_{S}$ and $\chi_{\rm I}$, may be precisely defined with respect to canonical ensemble averages:  

\begin{equation}
\chi_T = \frac{\partial M}{\partial H} =  \frac{1}{V}\frac{\partial}{\partial H}
\bigg(\sum_i m_i p_i\bigg),
\label{equation1}
\end{equation}

\begin{equation}
\chi_S = \chi_T - \frac{\mu_0 V T(\partial M/\partial T)^2}{C_{H}},
\label{equation2}
\end{equation}
\begin{equation}
\chi_{\rm I} =\frac{1}{V}\sum_i \left(\frac{\partial m_i}{\partial H}\right)\, p_i,
\label{equation3}
\end{equation}
where the sum is over eigenstates $i$ of the Hamiltonian (which are not generally simple spin states), $H$ is the applied field, $M$ is the equilibrium magnetization,  $V$ is the volume, $m_i = - \partial E_i /\partial B$ (with $B=\mu_0 H$) is the magnetic moment of eigenstate $i$,  $C_H$ is the magnetic heat capacity at constant applied field, $p_i = e^{-E_i/k_{\rm B} T}/Q$ is a Boltzmann population and $Q=\sum_{i}e^{-E_i/k_{\rm B} T}$ is the partition function, with $k{\rm_B}$ Boltzmann's constant. Note that the quantum adiabatic susceptibility $\chi_{\rm I}$ is not necessarily equal to the thermodynamic adiabatic susceptibility $\chi_{\rm S}$. Also,  $\chi_{\rm I}$ cannot be expressed as a second derivative of the free energy, so is not a thermodynamic property. It has been proved that $\chi_{T}\ge \chi_{S}\ge \chi_{\rm I} \ge 0$~\cite{Wilcox1968}.

In experiment, the frequency dependent susceptibility $\chi(\omega)$ generally measures $\partial M/\partial H = \chi_T$ as $\omega\rightarrow 0$. A purely real response $\chi(\omega) = \chi'(\omega)$ at high frequency could equate to 
$\chi_{S}$ if the only effect of finite frequency is to decouple the system from the heat bath~\cite{Casimir1938}, or it could equate to $\chi_{\rm I}$ if the state populations of the system remain equal to those that existed before the field perturbation was applied. In the latter case, if the fixed values $p_i$ are not equilibrium populations for all $H(t)$, then the response of the system is non-ergodic. The experiment measures a time ($t$) average, $\partial M(t)/\partial H(t)$, that is equal to $\chi_{\rm I}$, but is not equal to the ensemble averaged $\partial M /\partial H = \chi_T$. However, $\chi_{\rm I}$ can still be calculated  by a different average over the canonical ensemble, as given in Eq. 3.

\begin{figure}[t]
\includegraphics[width=9cm]{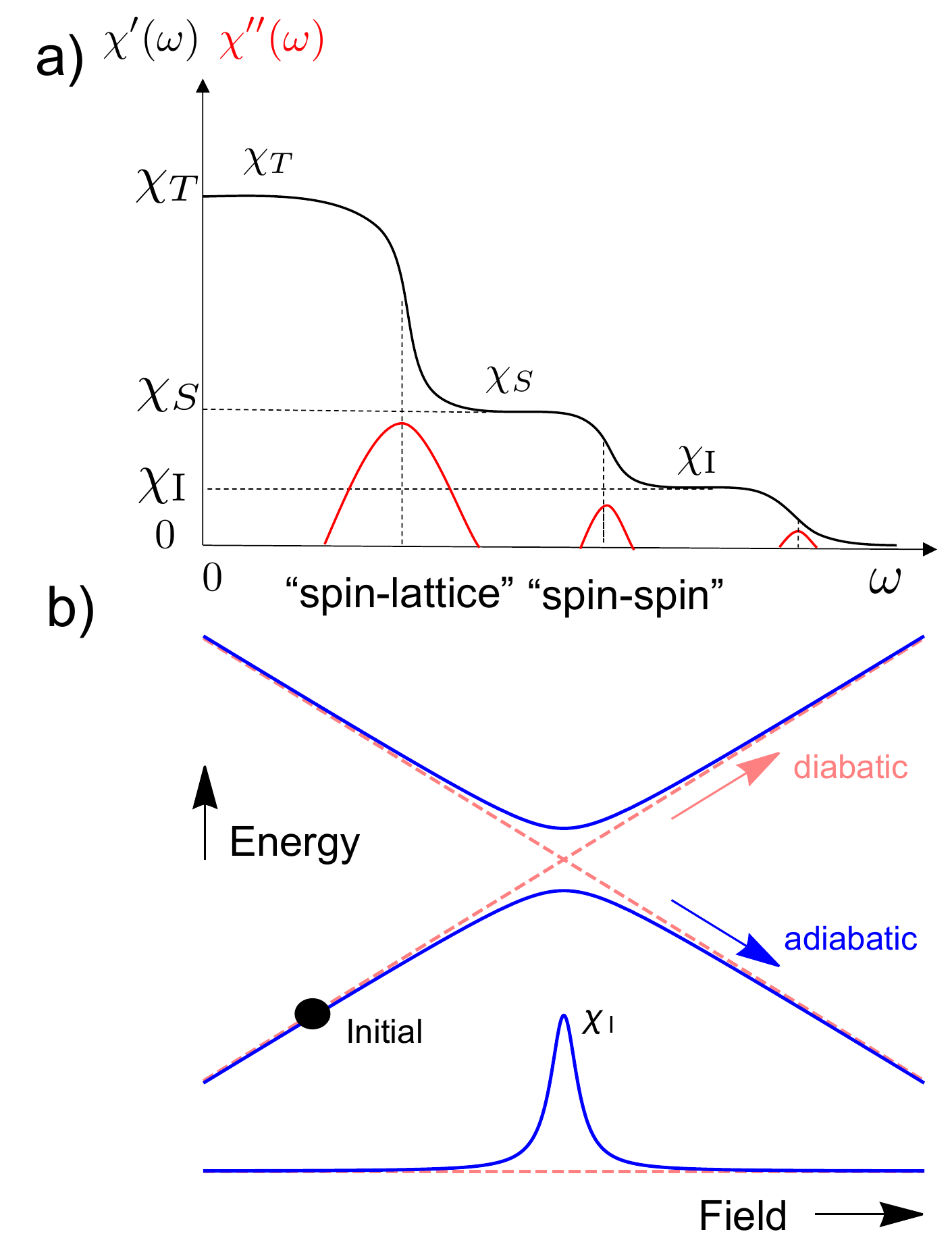}
\caption{(a) A schematic showing the frequency dependent decoupling of the real (black) and imaginary (red) parts of the magnetic susceptibility in a system with well-separated spin-lattice and spin-spin relaxation times. One can expect three plateaus of purely real response corresponding to $\chi_T$, $\chi_S$ and $\chi_{\rm I}$ respectively. (b) Avoided level crossing (upper blue curves). In this paper we are solely interested in adiabatic evolution (blue). This is illustrated for an initial state on the lower branch, as indicated by a black circle.  Also illustrated is diabatic evolution (pink) that could occur if there were a Landau-Zener crossover at frequencies greater than those studied here. The lower curves (same colour code) indicate the corresponding isolated susceptibilities for the ensemble of two state systems treated in Section II, where the adiabatic case yields a peak in $\chi_{\rm I}$ while the diabatic case gives zero (see Fig. 1a).
}
\label{one}
\end{figure}

To see how this may come about in practice, in Fig.~\ref{one}a we describe an idealized spin system where the driving period $\tau = 2\pi/\omega$ is compared to well-separated spin-lattice ($\tau_1$) and spin-spin ($\tau_{\rm 2} \ll \tau_1$) relaxation times~\footnote{T. Moriya has drawn a very similar figure: T. Moriya, Busseryon Kenkyu {\bf 3}, 332 (1958), Fig. 6.}. At very low frequency, the magnetic system will remain in thermal equilibrium throughout the field cycle, giving $\chi_T$ as the real response. As $\omega$ is increased until $\tau \ll \tau_1$, equilibrium with the lattice and heat bath are lost, but spin-spin interactions retain thermal equilibrium between spins, giving an adiabatic response, $\chi(\omega) = \chi_S$~\cite{Casimir1938}. If the drive frequency is further increased until $\tau \ll \tau_2$ then equilibrium between spins is lost and the perturbing field acts on the state populations that existed before the perturbation was applied, so $\chi(\omega) = \chi_{\rm I}$. Hence, with $\tau_1$ and $\tau_2$ well defined and well separated, the susceptibility $\chi(\omega)$ takes the form of a series of decreasing plateaus (Fig.~\ref{one}a), reminiscent of a dielectric response. As $\omega$ is increased, the susceptibility on each plateau  -- $\chi_T$, $\chi_S$ and $\chi_{\rm I}$ respectively  -- can be calculated in a quasi-static limit (Eq.~1-3). 

In this paper, we will be particularly interested in avoided level crossings (Fig. 1b) where the curvature of the state energies with field allows a finite $\chi_{\rm I}$ according to Eq. 3. In this context, it is important to stress that the isolated response is, by definition, {\it adiabatic} in the quantum mechanical sense~\cite{LL}.  The Landau-Zener effect~\cite{Rubbmark} (a dynamical effect) is therefore not relevant to this paper except insofar as it could imply a further crossover, with increasing frequency, from the case where the crossing is traversed adiabatically, to the case where it is traversed diabatically (see Fig. 1b). In view of the subsequent discussion, this would drive the susceptibility to zero, a feature that we have illustrated in  Fig. 1a. An experimental example of a Landau-Zener crossover in a rare earth complex is given in Ref. \cite{Wolfgang}.

The response of real magnetic systems can be far more complicated than implied by the simplified picture of Fig. 1a, but there will always be a gradual crossover, with increasing frequency, from a scenario in which state populations change on the time scale of the field cycle, to one in which they do not.  The low frequency regime can be treated by a master equation approach (see, for example, Refs. \cite{Pirc1974, Leuenberger2000}), which accounts for state population changes, while the high frequency regime can be treated by a Kubo-type linear response approach~\cite{Kubo1957}, which assumes fixed populations. Indeed, neutron scattering, which probes the response at relatively high frequencies, is very successfully treated by the latter approach {\cite{ML1968}}.

In contrast to $\chi_T$, which irretrievably mixes field-induced changes in quantum states with field-induced changes in state populations, 
$\chi_{\rm I}$ may be viewed as a more direct measure of quantum spectra. It is further a general measure of the superposition of quasi-classical spin states. Thus, if an energy eigenstate is say a pure `spin up' state, it will have zero isolated susceptibility because the magnetic moment is field independent (see Eq. 3). A finite $\chi_{\rm I}$ can, however, be observed if spin up and spin down are superposed, which allows the magnetic moment to evolve with field. This   
is confirmed by writing $\chi_{\rm I}$ in the equivalent form~\cite{Pirc1974}: 
\begin{equation}
\chi_{\rm I} 
=\mu_0 (V Q)^{-1} \sum_{i,j~(E_j \ne E_i)} \frac{e^{-\beta E_i} - e^{-\beta E_j}}{E_j-E_i} \left|\langle i\left|\hat{\mu}\right| j\rangle\right|^2,
\end{equation}
where $\hat{\mu}$ is the magnetic moment operator and $\beta = 1/k{\rm_B}T$. This contains finite matrix elements only if different energy eigenstates $i,j$ contain both spin up and spin down components, as occurs at avoided level crossings, for example. 

In this paper we demonstrate a particularly simple magnetic system in which the isolated susceptibility can be measured and analysed. The paper first considers relevant theory and then describes our experimental results. In Sec. II we solve the statistical mechanics of a simple two state paramagnetic system with an idealised spin Hamiltonian, designed to emphasise  differences between the three susceptibilities (Eqs. \ref{equation1}-\ref{equation3}) and to highlight the connection between isolated susceptibility and state concurrence. Sec. III shows how a variant of this Hamiltonian may be realised by approximating the hyperfine Hamiltonian of dilute spin ice. In particular, there is predicted a strong isolated response and perfect state concurrence at avoided level crossings in finite field. Sec. IV then describes our experiments that confirm these predictions at $T \ge 2$ K, where the system is prepared in a state of thermal equilibrium. Sec. V describes analogous low temperature experiments, down to $T = 76$ mK, that reveal how, when the system cannot be brought to equilibrium, isolated susceptibility can be used as a sensitive probe of the non-equilibrium state populations. Conclusions are subsequently drawn in Sec. VI. 

\section{II A two state system}

We start by considering a simple two-state system in which the difference between the three susceptibilities of Eqs. 1-3 may be made explicit. The system consists of an ensemble of non interacting spins, each with $S = 1/2$ and Hamiltonian
\begin{equation}\label{ham}
\hat{H} =  2 \mu B \hat{S}^z + 2 \Delta \hat{S}^x. 
\end{equation}
Here, $\mu$ is the size of the magnetic moment of the pure spin up or down states, $\Delta$ is a perturbation and the spin ($\hat S$) operators are defined to be dimensionless.
The first term in Eq. \ref{ham} is the Zeeman interaction and the second term mixes magnetic (`spin up' and `spin down') states.
This Hamiltonian is easily diagonalized, with eigenvalues characteristic of an avoided level crossing at $B=0$ (see Fig. 1b, blue lines):
\begin{equation}
E_{\pm} = \pm \mathcal{E}; ~~~~~~\mathcal{E} =  \sqrt{( \mu B)^2 + \Delta^2},
\end{equation}
and eigenstates: 
\begin{equation}
|\psi_\pm\rangle = \frac{  \Delta \left|\uparrow \right \rangle + (\mathcal{E} \mp \mu B)\left|\downarrow \right \rangle}
{\sqrt{(\pm \mathcal{E}-\mu B)^2 +\Delta^2}},
\end{equation}
characteristic of a gradual superposition of spin up and spin down as the crossing is approached.  The two quasi-classical states are completely superposed at the level crossing in zero field, but are not superposed far from that point.

To characterise this behaviour quantitatively, following Ref. ~\cite{Wooters1998} we may introduce the `concurrence' $\mathcal{C}$ of a state $|\psi\rangle$ as its overlap with its spin-reversed equivalent $|\tilde{\psi}\rangle$:
\begin{equation}
\mathcal{C}(\psi) = \left| \langle\psi | \tilde{\psi}\rangle \right|.
\end{equation}
This takes the value $\mathcal{C}=0$ for the non-superposed states $\left|\uparrow \right \rangle $ and $\left|\downarrow \right \rangle$ which are the eigenstates corresponding to $E_\pm$ in sufficiently strong fields $|B|$, and $\mathcal{C}=1$  for the fully superposed states $\psi_\pm = \left(\left|\uparrow \right \rangle \pm \left|\downarrow \right \rangle\right)/\sqrt{2}$, which are the eigenstates at the zero field avoided level crossing. It is then easy to show that the isolated susceptibility is a direct measure of concurrence as a function of field $\mathcal{C}(B)$, given by
\begin{equation}
\chi_{\rm I} = \frac{\mu_0}{V} \frac{\mu^2}{\Delta}  \sum_\pm \mp p_\pm \mathcal{C}^3,
\end{equation}
where $\mathcal{C}(B) = \Delta/\sqrt{\Delta^2 + B^2 \mu^2}$ and $p_\pm$ represent the normalized state populations. 
Using Boltzmann probabilities, this becomes
\begin{equation}
\chi_{\rm I} = \frac{\mu_0}{V} \frac{\mu^2\mathcal{C}^3}{\Delta}\,\tanh\left(\frac{\beta \Delta}{\mathcal{C}}\right)   \approx \mu_0 V^{-1} \beta\mu^2 \mathcal{C}(B)^2,
\end{equation}
where the right hand approximation is valid in the high temperature limit. The isolated susceptibility is then a Lorentzian function of field:   
\begin{equation}
\chi_{\rm I} = \frac{\mu_0 \mu^2}{V k_{\rm B} T} \times \frac{1}{1+ (\mu^2/\Delta^2) B^2}.
\end{equation}

With knowledge of the exact eigenstate energies as a function of field, all three susceptibilities of Eqs. 1-3 can therefore be calculated.  The isolated response is a maximum at $B = 0$ when $\mathcal{C} = 1$ and $\chi_{\rm I}(0) = \chi_S(0) =  \chi_T(0) = \mu_0 V^{-1} \beta\mu^2$. The closely spaced eigenstates of the system with energy $E_\pm=\pm \Delta$ have precisely zero magnetic moment at the avoided crossing in zero field, only gaining a moment through their mixing by the second order Zeeman effect as the field is applied. However with increasing field, the three susceptibilities diminish at different rates, such that $\mu_0 V^{-1} \beta\mu^2> \chi_{\rm T}(B) > \chi_{\rm S}(B) > \chi_{\rm I}(B)$.

The more complex Hamiltonian considered subsequently will retain many of the characteristics of this simple example, though some important differences of detail arise from the fact that in the more complex case, the avoided level crossings occur at finite applied field. 

\section{III Hyperfine Hamiltonian}

\begin{figure*}[t]
\centering{\resizebox{1 \hsize}{!}{\includegraphics{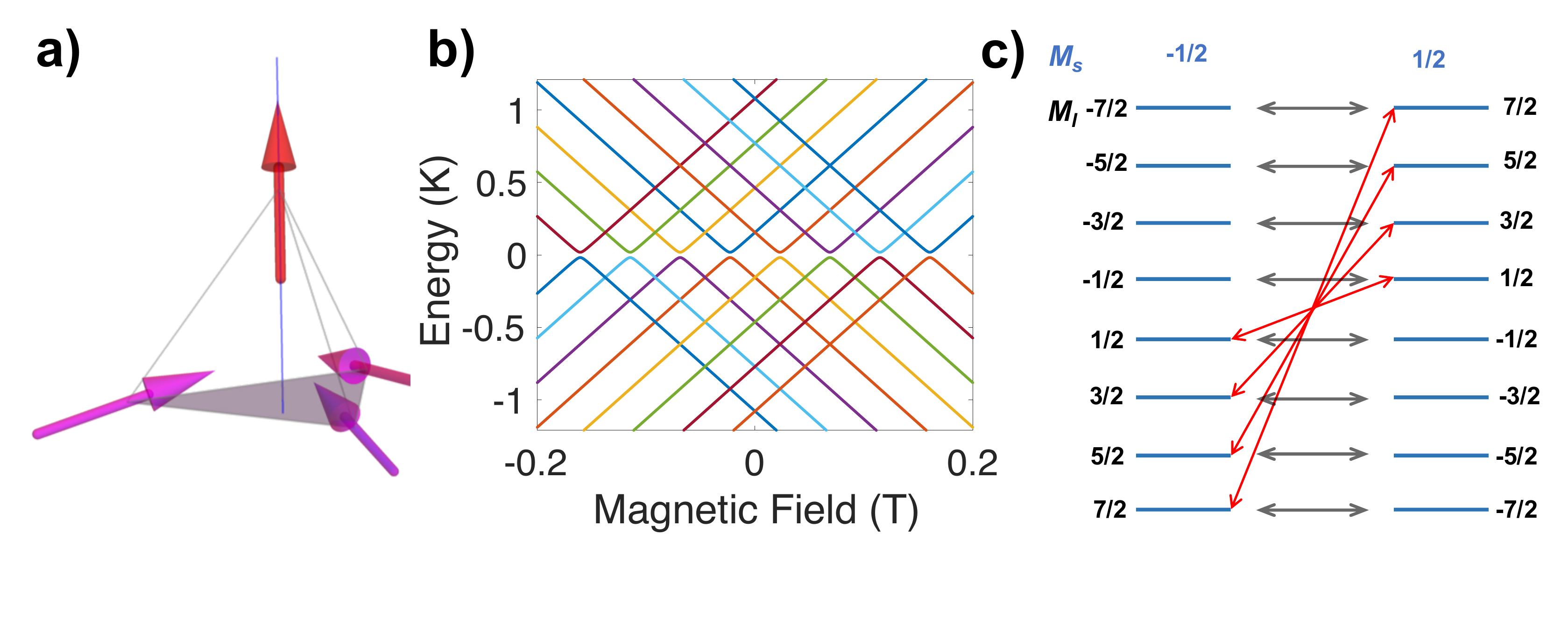}}}

\caption{({\it a}) The four Ising like $\langle 111 \rangle$ spin orientations of Ho$^{3+}$ with respect to the $[111]$ direction of the applied field (blue), with the basal plane shaded (apical spin in red, basal spins in magenta). In the dilute sample only one or zero magnetic sites are likely to be occupied by magnetic Ho$^{3+}$ in any given tetrahedron of the crystal structure. ({\it b}) Energy diagram of the hyperfine levels of an apical Ho$^{3+}$ ion as a function of magnetic field with an effective splitting of $\Delta/k_{\rm B} = 0.013$ K, showing direct and avoided level crossings. ({\it c}) The effective electronic spin-$1/2$ Ho$^{3+}$ ion has eight Zeeman split levels which are degenerate in zero field (gray arrows). As the frequency is increased only four transitions (red arrows) are allowed, such that only the electron spin reverses. When brought into resonance by applied field, superposed $\pm m_S$ states have finite isolated susceptibility.}
\label{two}
\end{figure*}

A model Hamiltonian similar to that considered may be realized experimentally in the dilute limit of an Ising-like spin system, such as very dilute samples of spin ice,  Y$_{1.9975}$Ho$_{0.0025}$Ti$_2$O$_7$, studied here. Fig. \ref{two}a illustrates the spin ice geometry where every fourth (apical) spin is parallel or antiparallel to the field, while the other three (basal) spins have only a small parallel or antiparallel component. We will be interested mainly in the response of the apical spin, but the three basal spins also need to be considered as they provide an important correction.

The non-Kramers holmium (${\rm Ho}^{3+}$) ion with nuclear spin  $I = 7/2$ would be expected to afford a weakly `split' electronic spin doublet with effective spin $S = 1/2$~\cite{AandB} and $z=\langle111\rangle$ quantization axis ensured by the large trigonal crystal field~\cite{Rosenkranz2000}. For an ideal non-Kramers  doublet of this sort, neglecting the quadrupole term, the hyperfine Hamiltonian may be written 
\begin{equation}
H_{\rm hyper} = A \hat{S}^z \hat{I}^z  ~~~~~~~~~~
A = g_\parallel \left(\frac{A_J}{g_J}\right)
\end{equation}
where 
%
$A_J/g_J$ should be essentially the same for all Ho salts (note that Ho has a single isotope).
 For Ho$^{3+}$ in the spin ice environment, the parallel g-factor is $g_\parallel\approx 19$ (see below) and the hyperfine parameter is $A/k_{\rm B}\approx 0.3$ K. If higher states are relevant, they can induce a further transverse term $A' (\hat{S}^x \hat{I}^x + \hat{S}^y \hat{I}^y)$ with $A' = g_\perp \left(\frac{A_J}{g_J}\right)$, but here, any perpendicular g-factor $g_\perp$ is certainly extremely small~\cite{Rosenkranz2000} and is henceforth assumed to be zero. More importantly, Abragam and Bleaney~\cite{AandB} recommend the addition of a term $2 \Delta (\hat{S}^x + \hat{S}^y)$ with a distribution of parameters $\Delta$ arising from local strains. We can suppress the $y$-term without loss of generality, and 
including the Zeeman term for the field $B$ applied parallel to $[111]$, the effective Hamiltonian for the apical spins becomes: 
\begin{equation}\label{apical}
\hat{H}_{\rm apical} = 2 A \hat{I}^z \hat{S}^z+ 2 \mu B \hat{S}^z + 2 \Delta \hat{S}^x,
\end{equation}
where $\mu = (1/2) g_\parallel \mu_{\rm B} \approx 10~\mu_{\rm B}$. A similar Hamiltonian may be constructed for the basal spins:
\begin{equation}\label{basal}
\hat{H}_{\rm basal} = 2 A \hat{I}^z \hat{S}^z+ 2 (\mu/3) B \hat{S}^z + 2 \Delta  \hat{S}^x, 
\end{equation}
The factor (1/3) in this equation arises from the angle $\arccos(1/3)$ that the basal spins subtend with the applied field.

Taking $\hat{H}_{\rm apical}$ as an example, these Hamiltonians may be represented in the basis of states $|m_S\rangle |m_I\rangle$, where the only off diagonal terms are those with $m_I - m_I' = 0$ and $m_S-m_S' = \pm1$. Here $m_S = \pm 1/2$ and $m_I = \pm 7/2, \pm 5/2, \pm 3/2, \pm 1/2$. Hence the Hamiltonian can be separated into a series of $2\times 2$
blocks of the type
\begin{equation}\label{real}
\hat{H}' =
\begin{bmatrix}
 A m_I +\mu B  & \Delta  \\
 \Delta & -A m_I -\mu B \\
 \end{bmatrix},
\end{equation}
one for each value of $m_I$. The Hamiltonian is thus easily diagonalized, as above, with eigenvalues:
\begin{equation}\label{emi}
E_{m_I,\pm} = \pm \sqrt{(A m_I + \mu B)^2 + \Delta^2}.
\end{equation}
Hence energies and magnetic moments of the Hamiltonian Eq. \ref{real} map to those of Eq. \ref{ham} - 7, with $\mu B$ replaced by  $A m_I + \mu B$: the nuclear spin acts as an effective field that adds to the applied field. It is then straightforward to evaluate the partition function corresponding to  Eq.~\ref{emi} and then derive the susceptibilities using Eqs. 1-3 with the magnetic moment defined as $m_i = -\partial E_i/\partial B$. 

The consequent energy diagram for the apical spins as a function of magnetic field is shown in Fig. \ref{two}b, illustrating the field dependent direct and avoided level crossings. 
There are two degenerate ladders of energy levels, each corresponding to  $m_S = \pm 1/2$, with $m_I$ defining the rungs of the ladder, as is depicted in Fig. \ref{two}c.  
Starting from the zero field as depicted in the figure, 
as a magnetic field is applied, and neglecting its weak coupling with the nuclear moment, the two sets of states shift by the electronic first order Zeeman energy difference. This gives, in principle, eight values of the applied field where energy level crossings occur. 
\begin{figure}[t]
\includegraphics[width=8cm,height=7cm]{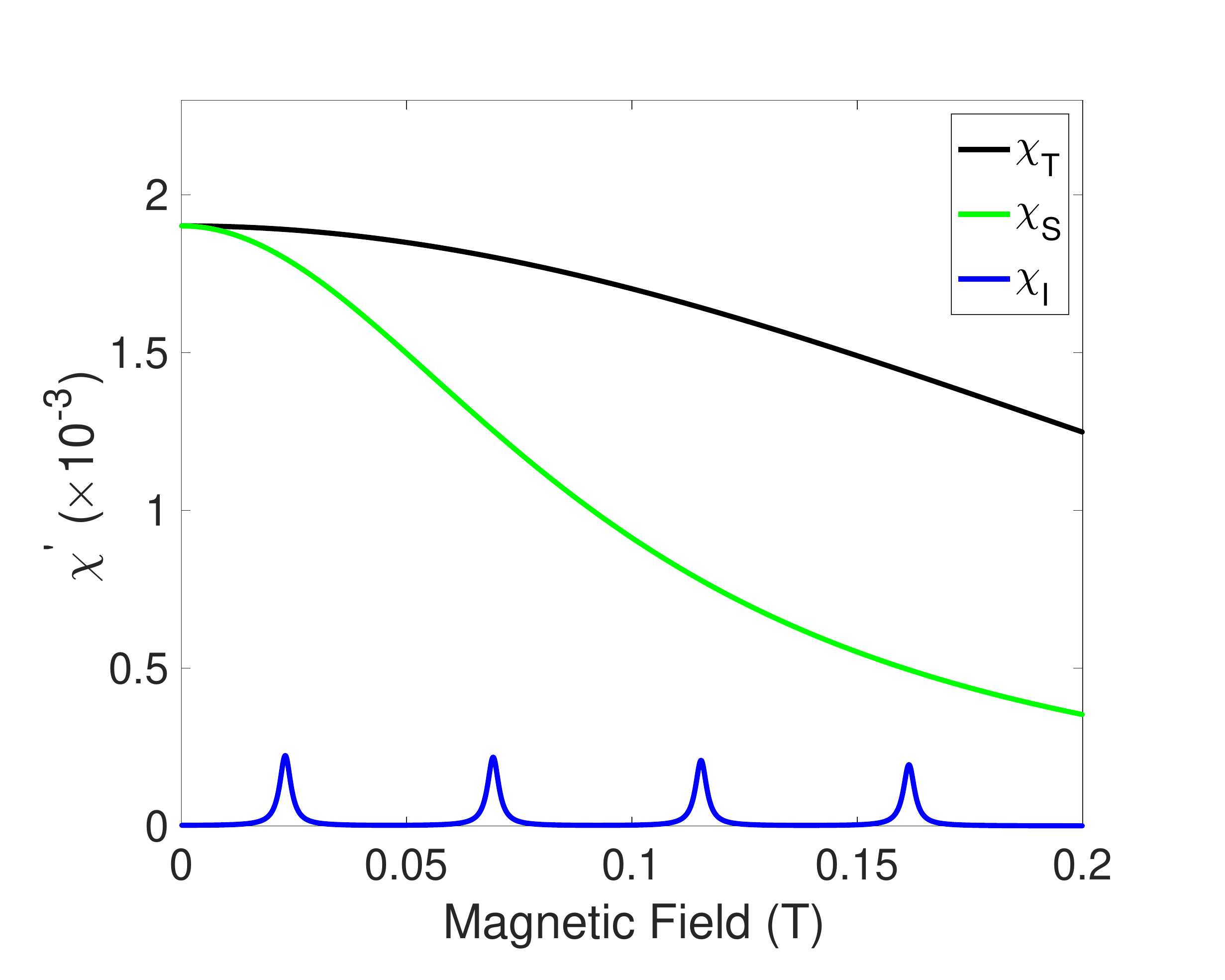}
\caption{Calculated values of the three susceptibilities (Eqs.~\ref{equation1}-\ref{equation3}) at $T = 2.1$ K for the apical spin Hamiltonian Eq.~\ref{apical}, with spin concentration equal to that of the sample studied.}
\label{three}
\end{figure}
However, the isolated susceptibility $\chi_{\rm I}(H)$  strongly peaks only at the subset of four avoided level crossings where the nuclear spin state does not change (see Fig. \ref{two}c). At these points, the `up' and `down' electronic spin states,
$\left|\uparrow \right \rangle$ and $\left|\downarrow \right \rangle$, respectively, are fully superposed and show complete concurrence, $\mathcal{C} =1$, while off-resonance, the spin states tend towards simple spin up or down with $\mathcal{C} = 0$ (see Eq. 7,8). However, in contrast to the simpler case considered above,  even at the special points of resonance, $\chi_{\rm I}$ is suppressed by a significant factor with respect to $\chi_T$ (see Fig.~\ref{three}). Physically, the reason is that, at a given avoided crossing, the isolated susceptibility is only finite for electronic states associated with a single nuclear spin state, a small fraction ($\sim1/(2I + 1)$ at high temperature) of the total available. One might say that the field-driven system is ergodic for one nuclear spin state, but non-ergodic for the rest, in contrast to the simpler case of Eq. 9, where the system is fully ergodic at the $B = 0$ resonance and $\chi_T = \chi_{\rm I}$ as envisaged by Kubo~\cite{Kubo1957}.

Considering just the isolated susceptibility, we refer to Fig. \ref{three}, and label the four isolated susceptibility peaks (left-right) as 1-4. A similar calculation (not shown) may be carried out for the basal spin Hamiltonian, Eq. \ref{basal}. The basal spins are found to contribute a small peak coincident with 2 and another small peak at three times that field (to the right of 4 in Fig. \ref{three}). The reason these features are relatively small is that the susceptibility scales as the moment squared and the projected moment of the basal spins is $1/3$ that of the apical ones, but they have three times the population so the peaks are only one third $(= 3 \times (1/3)^2)$ the height of the apical spin peaks. The basal spins therefore represent a small correction to peak 2 and a small rising background correction to the right of peak 4.

\begin{figure}[t]
\includegraphics[width=8cm,height=7cm]{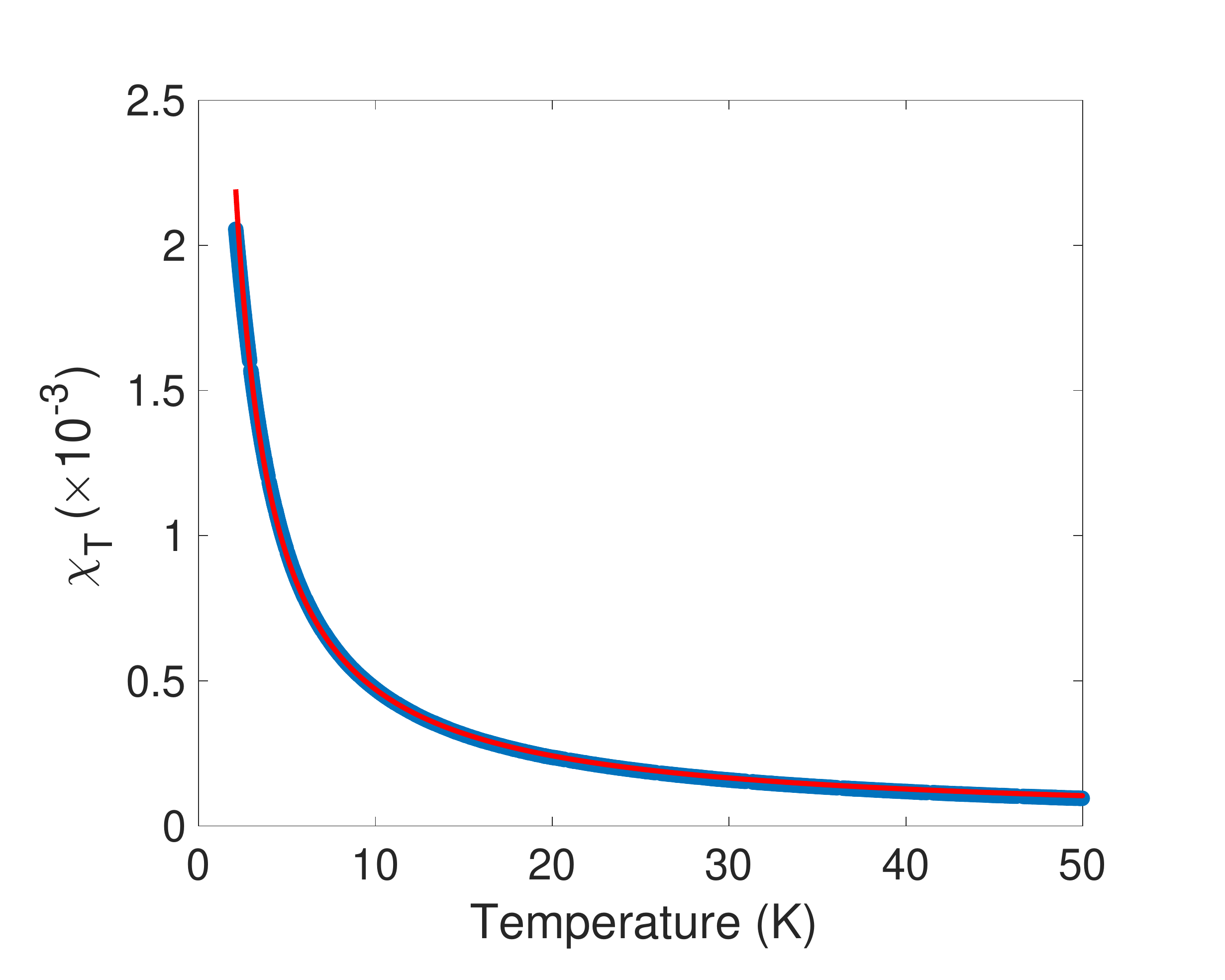}
\caption{Temperature dependence of the d.c. field cooled susceptibility of 
Y$_{1.9975}$Ho$_{0.0025}$Ti$_2$O$_7$
measured along the $[111]$ axis (blue points) compared to a Curie law fit summed with a small temperature independent component (red line).}
\label{four_new}
\end{figure}
More generally, a strongly structured nuclear-electronic response is indeed a long established behavior~\cite{Hufner1964} of dilute Ho$^{3+}$ ions in crystals, which have been impressively analysed with master equation-based approaches in several works \cite{Giraud2003,Bertaina2006,Graf2007,Barbara2008,Johnson2012}. In general, various transitions are possible depending upon coupling with the local environment of the effective spins, which generates a combination of intrinsic and interaction-induced direct and avoided level crossings, respectively. Our Hamiltonian, Eq. \ref{real} describes a highly simplified limiting case where the response is mapped on to the quasi-static isolated susceptibility, Eq. \ref{equation3}.  

\section{IV Experiment}

A single crystal of Y$_{1.9975}$Ho$_{0.0025}$Ti$_2$O$_7$
was grown by the optical floating-zone technique~\cite{Prabhak}. 
It was aligned with the applied field along the cubic $[111]$ axis. a.c. and d.c. susceptibility measurements were made at $T \ge 2$ K using a Quantum Design Physical Property Measurement System and at $T < 2$ K using a low-temperature SQUID magnetometer developed at the Institut N\'{e}el in Grenoble~\cite{paulsen_book}.
\begin{figure}[t]
\includegraphics[width=9.5cm,height=7.5cm]{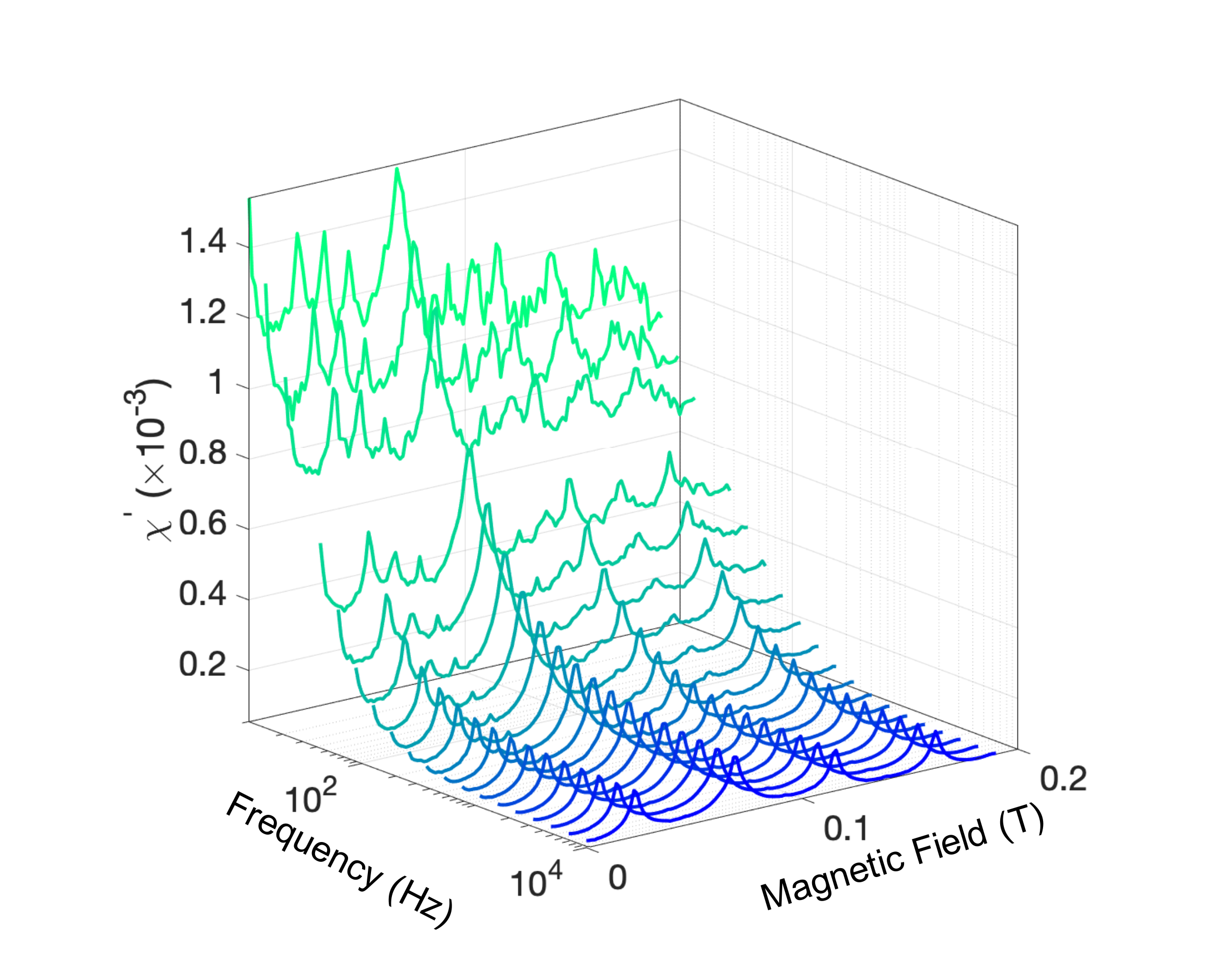}
\caption{Frequency and magnetic field dependence of the real part of the susceptibility measured at 2.1 K with an a.c. amplitude of 0.2 mT, demonstrating a gradual reduction to four peaks as the frequency is increased.}
\label{fivea}
\end{figure}

To confirm the stoichiometry $x$ and the paramagnetic approximation, the bulk susceptibility was measured, after cooling in a field of 0.1 T. The data were fitted to the sum of a Curie law $\chi_T = C/T$ and a very small temperature-independent component. The Curie constant $C$ was specified for density $x= 0.0025$ while the  g-factor $g_\parallel$ entering into the theoretical expression for $C$~\cite{Bovo_2013} was treated as a fitting parameter. This gave $g_\parallel = 19.0$, which we use in subsequent analysis. The excellent fit shown in Fig.~4 confirms that the nominal $x = 0.0025$ is accurate. A splitting between field cooled and zero field cooled susceptibility (not shown) was observed below $T = 3.6$ K. This shows that the spins are already falling out of equilibrium on the time scale of this `static' measurement. It is consistent with our observations of the frequency dependent susceptibility, as described below.

\begin{figure*}[t]

\centering{\resizebox{.9 \hsize}{!}{\includegraphics{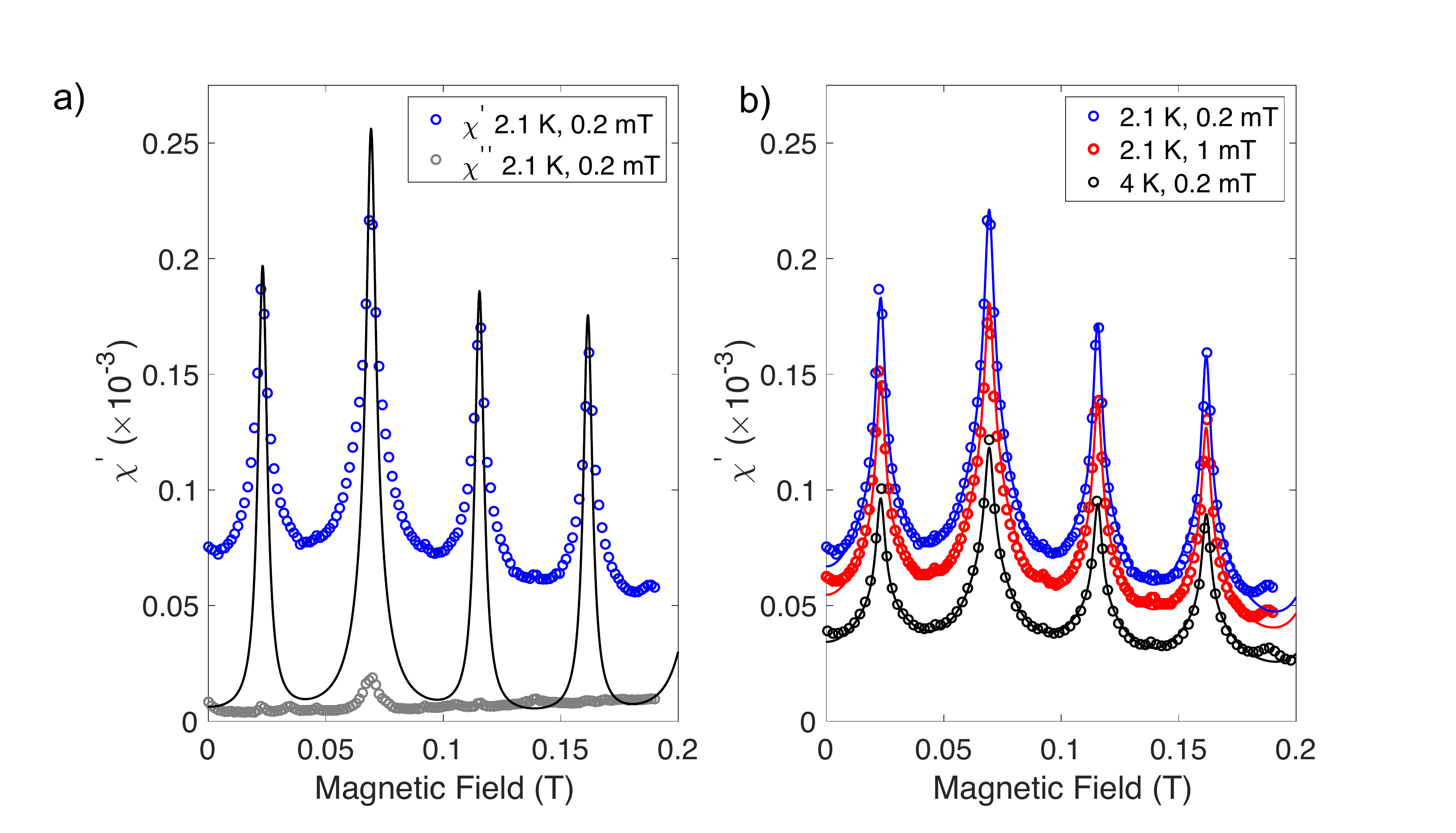}}}
\caption{{\it a.} The real (blue points) and imaginary (grey points) part of the measured susceptibility at 2.1 K, 10 kHz  and probe field 0.2 mT  compared to the theory, with a single value of    $\Delta/k_{\rm B}$ = 0.015 K (black line). {\it b}. Experiment versus theory with an empirical distribution of $\Delta$ (Eq. 18). The fitted parameters are  $f_1 = 0.511(6), f_2 = 0.352(3)$ for 2.1 K, 0.2 mT,  $f_1 = 0.455(6), f_2 = 0.312(3)$ for 2.1 K, 1.0 mT; $f_1 = 0.52(1), f_2 = 0.335(4)$ for 4 K, 0.2 mT. These parameters affect the peak shape, not peak heights and are expect to evolve with probe field, but not significantly with temperature. }
\label{figfiveb}
\end{figure*}

The a.c. susceptibility $\chi'(\omega)$ measured at $2.1$~K (with a probe field of $0.2$~mT) is shown in Fig.~\ref{fivea}. At low frequency and zero field $\chi'(\omega)$ approaches $\chi_T$ of Fig. \ref{four_new}, and while it still falls short by some 25\% at the lowest applied frequency, the `$\chi_T$' plateau of Fig. 1a  can safely be presumed to exist at lower frequencies than those applied here. In finite field, multiple peaks evolve with increasing frequency to become four distinct peaks by 10 kHz. The spectrum and amplitude of these four peaks correspond closely to the isolated susceptibility of Fig. \ref{three}, and we may therefore identify the distinctive plateau in their frequency evolution with the `$\chi_{\rm I}$' plateau of Fig. 1a.  As anticipated, the high frequency susceptibility on the peaks is essentially a real response  with any imaginary component being only a few percent of the real part (see Fig. 6a). At intermediate frequencies the `$\chi_S$' plateau of Fig.1a is not clearly resolved at any value of applied field, presumably because spin-spin and spin-lattice times are not sharply defined or separated in this system. The multi-peak structure in the intermediate frequency range is similar to that described for other Ho systems~\cite{Bertaina2006}. 

It is clear from Fig. 5 that transitions at high frequency are of the type $\left|1/2 \right \rangle|m_I\rangle \to \left|-1/2 \right\rangle|m_I\rangle$, consistent with our effective Hamiltonian for spins parallel to the field. 
Also, as well as showing the four peaks associated with the apical spins, the experimental spectra evidence the expected additional small component on peak 2 from the basal spins: that is peak 2 is approximately 1/3 higher than the other peaks. There is also what seems, at first sight, like an unexpected `nonzero background' that fills in the gaps between the peaks.  However the fact it is mainly real (see Fig. 6a) identifies it as part of the isolated response itself.

We now consider how these data can be fitted quantitatively. The theoretical isolated susceptibility is found by summing the susceptibilities $\chi _{\rm I}$ of the apical and basal spins calculated as described above, using Eq.~\ref{equation3}, with Boltzmann probabilities for the nuclear-electronic spin states and the hyperfine parameter refined to $A/k_{\rm B} = 0.2945$ K by fitting the experimental peak positions.

With this, the fits have no adjustable parameters except those connected with the distribution of $\Delta$, which  
can only be estimated empirically~\cite{AandB}. Fig. 6a compares the experimental data with a single value of $\Delta/k_{\rm B} = 0.015$ K. The peaks are quite well described, but the regions between them are underestimated. The broadened bases of the peaks suggest a contribution to the distribution from $\Delta/k_{\rm B}\approx 0.1$ K. There is also the possibility of some ions  experiencing a very small $\Delta$ which raises a complication. For sufficiently small values of $\Delta$, the apparent response in an a.c. susceptibility experiment with finite probe field will approach zero, because of the nonlinearity (and eventual saturation) of the magnetic moment with field. Our experiment used a probe field of 0.2 mT from which we can estimate $\Delta/k_{\rm B} < 0.002$ K as the point beyond which the response will be suppressed.

To capture these properties in an empirical, parameterized, distribution, we consider one consisting of three delta functions (one at zero, and one each at the lower and upper values of $\Delta$ discussed above):
\begin{equation}\label{prob}
P(\Delta) = f_0 \delta(\Delta) + f_1 \delta(\Delta - 0.015~{\rm K}) +  f_2 \delta(\Delta - 0.1~{\rm K}),
\end{equation}
(where we have suppressed factors of Boltzmann's constant for clarity).
The delta function at zero does not contribute any response, but is relevant through the normalisation condition on the frequencies: $\sum_i f_i = 1$. 
With this, we have two independent parameters to fit the data: $f_1$ and $f_2$.

Fig. 6b compares fits to the data taken either at the same temperature (2.1 K) with a larger probe field (1 mT) or at a larger temperature (4 K) with the same probe field (0.2 mT). For the larger probe field, as would be anticipated, the `invisible' part of the response (represented by $f_0$) is increased as more of the nonlinear response is sampled over the field cycle. As expected, the susceptibility derived at 1 mT falls below that of the 0.2 mT measurement and $f_0 = 1-f_1-f_2$ derived from the fit was found to increase accordingly from 0.13 at 0.2 mT to 0.23 at 1 mT. Changing temperature, on the other hand yields parameters in close agreement with those found at $T=2$ K, again as expected. 

Referring to Fig. 6b, the apical response is clearly very well fitted by the model Hamiltonian.  The basal response is slightly less well described, as a small imaginary component is detectable on peak 2. However this part of the response is far too small to warrant introducing extra parameters into the model. The fact that a quite accurate fit is already obtained using a simple empirical distribution of $\Delta$ points to the conclusion that the model is essentially correct.\\\
\section{V Low temperature behaviour}

\begin{figure}[t]
\includegraphics[width=8cm,height=7cm]{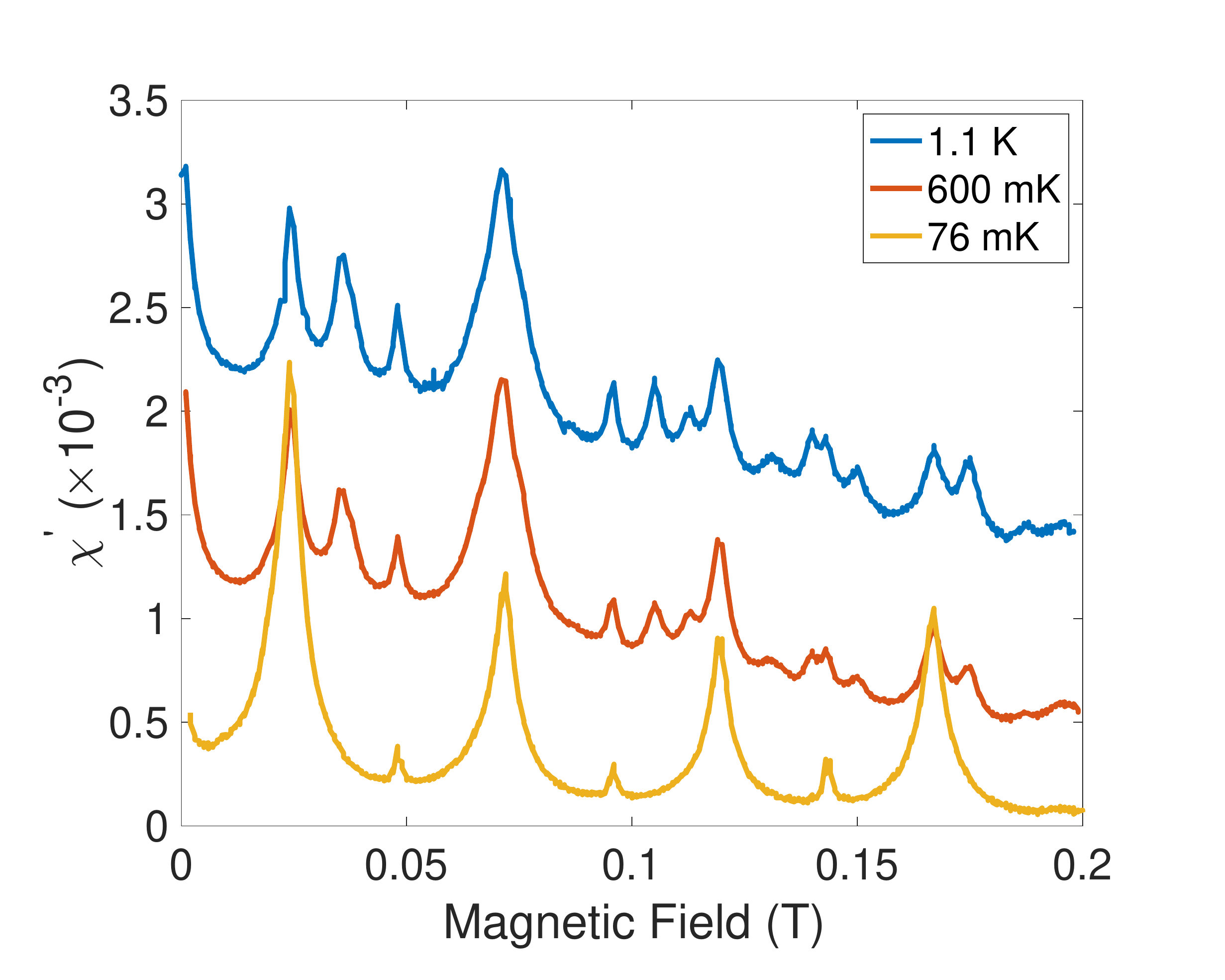}
\caption{The real part of the susceptibility at 11 Hz as a function of field and decreasing temperature, demonstrating that the isolated response is revealed by low temperature as well as high frequency, and that the distribution of peak intensities is no longer that of a Boltzmann distribution (contrast Fig. 6).}
\label{figsix}
\end{figure}

Because the isolated susceptibility separates population changes from state vector changes, the decline of peak intensity with increasing field (having allowed for the basal spin contribution on peak 2) is a measure of the decline in population with increasing energy, and hence a measure of the temperature. This gives an interesting method of directly investigating the effective temperatures reached when spin ice falls out of thermal equilibrium at $T < 0.6$ K \cite{Snyder2000}.  Fig.~\ref{figsix} shows the temperature dependence of $\chi'$ at a fixed frequency of 11 Hz as the dilute sample is cooled.  As the temperature is lowered to the base temperature of 76 mK, the four peaks indicative of isolated response again appear, but there are two features that mark these as reflecting non-equilibrium populations. First, at 76 mK, direct calculation shows that all peaks should have zero intensity if equilibrium is maintained, yet their observed intensities suggest a temperature of order 1 K. Second, peak 4 is now anomalously intense, which is a signature that the system, initially zero field cooled, does not fully re-equilibrate as the field is applied. Thus, referring to the energy level diagram of Fig. 2b, if the system retains the equilibrium populations of zero field, then the $I = 7/2$ resonance (a former ground state) will have the strongest intensity, not the weakest. We deduce that, in contrast to higher temperatures where the state populations are thermally equilibrated before the a.c. probe field is applied, at low temperature the system does not fully equilibrate in response to the changes in temperature and applied magnetic field that take place before the measurement.  Yet it is clear that Eq.~\ref{equation3} is valid, regardless of whether or not the $p_i$'s are Boltzmann populations, so observation of the isolated susceptibility can be used to measure how the actual $p_i$'s depend on energy. To analyse the details of this behavior theoretically is an interesting challenge that is beyond the scope of this paper.  We also remark that there is potential for the control of non-equilibrium state populations by pumping and then subsequent measurement by susceptometry.

\section{VI Conclusion}

In conclusion, although there has been much work on the susceptibility of isolated rare earth ions, including detailed master equation based analyses of experimental data  \cite{Giraud2003,Bertaina2006,Graf2007,Barbara2008,Johnson2012}, it appears that a simple reduction to a well defined isolated susceptibility has not been previously observed. Our observation of it in dilute spin ice  has been enabled by the unique local environment of the Ho$^{3+}$, which allows a very simple effective Hamiltonian to be enacted. 

Our results highlight the difference between the processes imagined when one formulates the thermodynamic adiabatic and quantum adiabatic (isolated) susceptibilities (Eqs. \ref{equation2} and \ref{equation3}) to describe the response of the system to a change in applied field, $B\rightarrow B+dB$. In the thermodynamic case, the process is thermodynamically reversible: the state of the system in the field $B+dB$ is a thermal equilibrium state. In the quantum case the process is only mechanically reversible: the system is out of thermal equilibrium in the field $B+dB$.  
This begs the question, if state populations cannot respond to a change in field, how could a Boltzmann population be prepared in the first place? The answer is that one needs the equilibrium state in a field $B$ to be prepared on some time scale that is sufficiently long for equilibrium to be established, but the perturbing field $dB$ to be applied on a time scale that is sufficiently short that changes in state population do not occur. Our experiments furnish examples that achieve this condition (here at higher temperature) as well as examples that do not (here at lower temperature). In the latter case our results indicate that a Boltzmann population cannot be prepared, and we would expect that the actual distribution (and hence response) will show a complicated history dependence: our initial measurements confirm this expectation, but to fully characterise this behaviour will be a major project. Here we confine ourselves to the conclusion that the isolated susceptibility may potentially be used to directly measure the 
non-equilibrium state populations. This indicates a promising avenue of research in the context of spin ice and other rare earth magnets, as it suggests a way to test non-equilibrium theories, the concept of effective temperature \cite{Cugliandolo}, and so on.

Finally, we have shown how the isolated susceptibility is a direct measure of concurrence $\mathcal{C}$ between spin states, with $\mathcal{C} = 1$ at the avoided level crossings. This infers that the apical spin population in our dilute spin ice sample shows complete, or nearly complete, concurrence at these special points. We are not aware of any other examples of an experimental measurement of state concurrence in a real magnetic system. Whether or not this ability to measure concurrence translates to more strongly interacting systems is an open question, but a strongly interacting system showing these effects may be afforded by bulk (concentrated) spin ice, where spin flipping associated with `monopole' excitations gives similar peaks in the high frequency susceptibility~\cite{Bovo, Schiffer}.


\begin{acknowledgments}
SRG and STB would like to thank EPSRC for funding, grant numbers EP/S016465/1 and  EP/S016554/1 respectively. EL and CP acknowledge financial support from ANR, France, Grant No. ANR-15-CE30-0004. SRG would also like to thank A. Armour for illuminating discussions.
\end{acknowledgments}

\bibliography{library2}

\end{document}